\documentclass[showpacs,preprintnumbers,twocolumn,amsmath,amssymb,prb,superscriptaddress]{revtex4-1}

\def\XXint#1#2#3{{\setbox0=\hbox{$#1{#2#3}{\int}$}
     \vcenter{\hbox{$#2#3$}}\kern-.5\wd0}}

\usepackage{graphicx}
\usepackage{subfigure}
\usepackage{bm}

\begin{document}

\title{Raman spectra of nanoparticles: elasticity theory-like approach for optical phonons}

\author{O.\ I.\ Utesov}
\email{utiosov@gmail.com}
\affiliation{National Research Center "Kurchatov Institute" B.P.\ Konstantinov Petersburg Nuclear Physics Institute, 188300 Gatchina, Russia}
\affiliation{St. Petersburg Academic University - Nanotechnology Research and Education Centre of the Russian Academy of Sciences, 194021 St. Petersburg, Russia}

\author{A.\ G. Yashenkin}
\affiliation{National Research Center "Kurchatov Institute" B.P.\ Konstantinov Petersburg Nuclear Physics Institute, 188300 Gatchina, Russia}
\affiliation{Department of Physics, Saint Petersburg State University, 198504 St.~Petersburg, Russia}

\author{S.~V.~Koniakhin}
\email{kon@mail.ioffe.ru}
\affiliation{Institut Pascal, PHOTON-N2, University Clermont Auvergne, CNRS, 4 avenue Blaise Pascal, 63178 Aubi\`{e}re Cedex, France}
\affiliation{St. Petersburg Academic University - Nanotechnology Research and Education Centre of the Russian Academy of Sciences, 194021 St. Petersburg, Russia}



\date{\today}

\begin{abstract}


A simple way to investigate theoretically the Raman spectra (RS) of nonpolar nanoparticles is proposed. For this aim we substitute the original lattice optical phonon eigenproblem by the continuous Klein-Fock-Gordon-like equation with Dirichlet boundary conditions. This approach provides the basis for the continuous description of optical phonons in the same manner how the elasticity theory describes the longwavelength acoustic phonons. Together with continuous reformulation of the bond polarization model it allows to calculate the RS of nanoparticles  without referring to their atomistic structure. It ensures the powerful tool for interpreting the experimental data, studying the effects of particle shape and their size distribution. We successfully fit recent experimental data on very small diamond and silicon particles, for which the commonly used phonon confinement model fails. The predictions of our theory are compared with recent results obtained within the dynamical matrix method - bond polarization model (DMM-BPM)  approach and an excellent agreement between them is found. The advantages of the present theory are its simplicity and the rapidity of calculations. We analyze how the RS are affected by the nanoparticle faceting and propose a simple power law for Raman peak position dependence on the facets number. The method of powder RS calculations is formulated and the limitations on the accuracy of our analysis are discussed.

\end{abstract}

\maketitle

\section{Introduction}

Nowadays, nanoparticle research is one of the most rapidly developing areas in condensed matter physics. Small particles are highly desired objects for material science~\cite{behler2009nanodiamond}, chemistry~\cite{xia2013nanoparticles}, biology and medicine~\cite{walling2009quantum,kim2016ultrasmall}, quantum computing~\cite{veldhorst2014addressable,andrich2017long}, photonics~\cite{riedel2017deterministic}, etc. The nanoparticles are manufactured in both ordered (photonic crystals and quantum dot massives) and random (powders and water suspensions) arrays, the latter require specific experimental techniques for their characterization. These techniques include X-Ray diffraction, atomic force microscopy, transmission electron microscopy, dynamic light scattering, Raman spectroscopy, etc. Among them, the optical nondestructive method of Raman spectroscopy plays very important role providing the information about nanoparticle size and phase composition.

While in bulk materials only the optical phonons from the Brillouin zone center  contribute to the main Raman peak, in nanocrystallites (e.g., diamond, Si, Ge, and GaAs) this peak is shifted to lower frequencies by several or even several dozens of reciprocal centimeters due to size quantization effect. Moreover, the peak becomes asymmetrically broadened. These phenomena can be used for nanoparticle characterization. 

In order to quantify these effects and to connect the peak position with nanoparticle size the phonon confinement model (PCM) is widely used~\cite{richter1981one,campbell1986effects}. This simple semiphenomenological model accounts for the finite size effects via the one-parametric Gaussian envelope function for the optical phonon amplitude within the nanoparticle. However, a deeper analysis of the PCM reveals a series of essential problems~\cite{zi1997comparison,meilakhs2016new}. Numerous attempts to modify the PCM did not resolve these problems but instead introduced more adjustable parameters \cite{osswald2009phonon,korepanov2017quantum,faraci2006modified,ke2011effect}. 

Very  recently, it was formulated~\cite{koniakhin2018novel} an alternative approach based on the combined use of dynamical matrix method and bond polarization model (DMM-BPM). In our opinion, this theory is based on more physical grounds than the PCM. Contrary to the latter, the DMM-BPM approach allows to interpret recent experimental data on nanodiamond powders very successfully~\cite{shenderova2011nitrogen,stehlik2015size,stehlik2016high}. The disadvantage of the DMM-BPM is that one of its constituents, the dynamical matrix method, requires diagonalization of huge $3N \times 3N$ matrices, where $N$ is the number of atoms in the crystallite. Although this general numerical method inspires the  quickly solvable analytical version, the latter misses some information about nanoparticle shape.

In the present paper we develope another approach to the Raman spectra (RS) analysis. The difficulty of description of \textit{optical} phonons is that the corresponding equations of motion do not possess straightforward continuous reformulation even in the bulk, unlike acoustic phonons \cite{landau1986theory,anselʹm1981introduction}. Our basic idea is to substitute the original atomistic optical phonon eigenproblem by an effective continuous media problem that has the spectrum coinciding with the spectrum of optical phonons in the longwavelength limit. It is performed in the same manner as the elasticity theory substitutes the ball-spring picture for acoustic vibrations. \textit{De facto}, we formulate the effective (isotropic and scalar) continuous theory for optical phonons similar to the elasticity theory for acoustic waves. We solve this problem (with proper boundary conditions) for finite size samples. This gives the possibility to calculate the RS and to compare them with the RS of original optical phonons (both experimental and theoretical).

We also apply the above theory in order to investigate the influence of shape of nanocrystallites on their Raman spectra. Studying the particle shape dependence of the main Raman peak we observe that the lack of knowledge concerning the shape leads to the uncertainty in size.

Finally, we formulate the method how to incorporate into our calculations the size distribution function on the most economical manner, when dealing with powder spectra.

Technically, we start from proper model for a crystal with diamond-type lattice namely from the linear chain with two atoms with equal masses and different spring rigidities in the unit cell~\cite{note2}. We formulate two separate continuous equations for acoustic and optical modes valid in the longwavelength limit. The optical mode is effectively described by the Klein-Fock-Gordon equation in the Euclidean space (EKFG). Analysis of a finite chain reveals the Dirichlet boundary conditions for our boundary value problem. Extending the treatment onto three dimensional case, we solve this eigenproblem analytically for simplest particle shapes (sphere and cube) and present numerical results for several important manifolds. 

We adapt our general theory in order to compare the developed approach with alternative theoretical and experimental descriptions of nanoparticles. For this purpose we reformulate the BPM in the continuous form suitable for RS construction within the EKFG theory. To show the potency and possible applications of our method we start from comparing theoretical EKFG and DMM-BPM spectra, and find an excellent agreement between the results. Furthermore, we successfully fit the experimental data on diamond~\cite{stehlik2015size} and silicon~\cite{expsi2011} powders. The only adjustable parameter of our theory is the phonon linewidth $\Gamma$.

Then we turn to the problem how the particle shape affects the Raman scattering. We study the RS of nine different (Platonic, Archimedian, spherical and elongated) shapes of the particles. We found empirical parabolic dependence of the main peak position on the inverse number of facets for Platonic and Archimedean solids and for sphere. However, due to very broad phonon linewidth observed experimentally \cite{yoshikawa1995raman}  the Raman spectra of these particles acquire almost universal form, so in practice the RS for various shaped nanoparticles collapse onto a single curve. 

We conclude that the deficit of information about the particle shape  provides the natural limitation for determining the particle size from the Raman scattering experiment with accuracy $\sim 10\%$. Additional information can be extracted from (presently unique) experiments on transmission electron microscopy (TEM) imaging  with atomic planes resolution\cite{dideikin2017rehybridization} and exploiting general concept of similarity between crystalline habitus of macroscopic crystals and nanoparticles. On the other hand, for crude rapid analysis of RS one can use the simplest analytically solvable particle shapes such as cube or sphere.

At last, combining the information obtained by means of DMM-BPM and EKFG approaches we obtain a scaling of RS for particles of different size. We propose a simple recipe how to construct the RS of arbitrary sized particles from the spectrum of a particle of a given size. Using this recipe we obtain powder RS from the single-particle RS. This method is valid for arbitrary size distribution function and does not require the recalculation of RS for every particle size.

The rest of the paper is organized as follows. In Sec.~\ref{SKFG} we derive the EKFG equation for optical phonons. In Sec.~\ref{SNano} we present the analytical solutions of this equation for two particular shapes of nanoparticles, namely for cube and sphere. We also connect material constants in EKFG equation and the constants of Keating model. In Sec.~\ref{SRS} we formulate the BPM in the continuous form. We compare predictions of our theory with PCM and DMM-BPM theoretical approaches and fit the most recent experimental data in Sec.~\ref{SCompar}. In Sec.~\ref{SShape} we evaluate numerically the RS for various Platonic and Archimedean solids and for the sphere, and collapse them onto a single curve. Sec.~\ref{SScale} elucidates the role of the size distribution function in our calculations of powder RS. Finally, in Sec.~\ref{SSum} we discuss obtained results and present the summary.        


\section{Klein-Fock-Gordon-like equation}
\label{SKFG}

In this Section we derive the effective \textit{continuous} equation for \textit{optical} phonons. Under certain approximations we decouple the finite-differences equations and obtain the continuous ones describing acoustic and optical branches of the spectrum. We generalize this approach onto three dimensions and formulate the boundary value problem with Dirichlet boundary conditions for optical mode.  

As the model system for nonpolar crystals (diamond, silicon, etc.) we consider the linear chain which consists of two atoms with equal masses and different rigidities of intra- and inter- cell springs, $k_1$ and $k_2$, respectively. The distance between the atoms within the unit cell is $a$ and between the atoms in neighboring unit cells is $b$, the lattice parameter being $a_0=a+b$. The second Newton law reads
\begin{eqnarray}
  \label{x1}
  m \ddot{x}_j &=& -k_1 (x_j -y_j)- k_2 (x_j - y_{j-1}), \\
  \label{y1}
  m \ddot{y}_j &=& -k_1 (y_j -x_j)- k_2 (y_j - x_{j+1}).
\end{eqnarray}
Introducing Fourier transform
\begin{eqnarray}
  x_j &=& x_q e^{i(q(a+b)j -\omega t)}, \\
  y_j &=& y_q e^{i(q[(a+b)j+a] -\omega t)},
\end{eqnarray}
one can write dispersion relation in the form
\begin{equation} \label{Spec1}
  \omega^4 - \frac{2(k_1+k_2)}{m}\omega^2+\frac{2k_1k_2(1-\cos{qa_0})}{m^2}=0,
\end{equation}
which yields
\begin{equation} \label{Spec2}
  \omega^2=\frac{k_1+k_2}{m} \left( 1 \pm \sqrt{1 - \frac{4 k_1 k_2}{(k_1+k_2)^2}\sin^2{\frac{qa_0}{2}}} \right),
\end{equation}
where ``$+$'' corresponds to the optical branch and ``$-$'' to the acoustic one. For longwavelength acoustic phonons we have:
\begin{equation} \label{SpecA}
  \omega^2_{ac} \approx \frac{k_1 k_2}{2m(k_1+k_2)}q^2 a_0^2, 
\end{equation} 
with the sound velocity given by
\begin{equation} \label{Sound}
  c = a_0 \sqrt{\frac{k_1 k_2}{2m(k_1+k_2)}}. 
\end{equation} 
For the optical branch near the Brillouin zone center we obtain
\begin{equation}
  \label{SpecO}
  \omega^2_{opt} \approx \frac{2(k_1+k_2)}{m}-\frac{k_1 k_2}{2m(k_1+k_2)}q^2 a_0^2.
\end{equation}
It is easy to recognize that $x_q \approx y_q$ for acoustic phonons whereas $x_q \approx -y_q$ for the optical ones. Bearing this in mind we rewrite Eqs.~ \eqref{x1} and \eqref{y1} in the form:
\begin{eqnarray}
  \label{X1}
  m (\ddot{x}_j+\ddot{y}_j) &=& -k_2 (x_j + y_j)+ k_2 (x_{j+1} + y_{j-1}), \\
  \label{Y1}
  m (\ddot{x}_j-\ddot{y}_j) &=& -2 (k_1+k_2) (x_j -y_j)+ k_2 (y_{i-1} - y_{j}) \nonumber   \\  && - k_2 (x_{i+1}-x_i). 
\end{eqnarray}
Now we introduce new (symmetrical and antisymmetrical) displacements $X=x+y$ and $Y=x-y$. In the longwavelength limit we can treat them as continuous functions of coordinate $z$ along the chain. Then Eqs.~\eqref{X1} and \eqref{Y1} can be rewritten up to the second order in $z$ derivatives as follows
\begin{eqnarray}
  \label{X2}
  m \ddot X &=& k_2 a_0 Y^\prime + \frac{k_2 a_0^2}{2} X^{\prime\prime}, \\
  \label{Y2}
  m \ddot{Y} &=& -2 (k_1+k_2) Y - k_2 a_0 X^\prime - \frac{k_2 a_0^2}{2} Y^{\prime\prime},
\end{eqnarray}
where $^\prime=\partial_z$. Neglecting slowly varying in time and space terms, $\ddot{Y}$ and $Y^{\prime\prime}$ (remember that $\partial_z \sim q \ll \pi/a_0$ ant thus these terms are proportional to $q^2$), for the gapless acoustic ($\omega \approx c q \ll \sqrt{k_{1,2}/m}$) branch Eq.~\eqref{Y2} yields 
\begin{equation}
  Y \approx -\frac{k_2 a_0 }{2(k_1+k_2)}X^\prime,
\end{equation}
and therefore
\begin{equation}
  \label{Y3}
  Y^\prime = -\frac{k_2 a_0 }{2(k_1+k_2)}X^{\prime\prime}.
\end{equation}
Plugging Eq.~\eqref{Y3} into Eq.~\eqref{X2} we obtain
\begin{equation}
  \label{Acoust}
  \ddot{X} = \frac{k_1 k_2 a_0^2}{2m(k_1+k_2)} X^{\prime \prime},
\end{equation}
which is the acoustic wave equation. 

Similarly, for a gapped ($\omega \approx \sqrt{2(k_1+k_2)/m}$) mode from Eq.~\eqref{X2} one finds
\begin{equation}
  X^\prime= -\frac{k_2 a_0 }{m \omega^2} Y^{\prime\prime} \approx -\frac{k_2 a_0}{2(k_1+k_2)} Y^{\prime\prime},
\end{equation}
and we finally get the continuous equation for the optical mode
\begin{equation}
  \ddot{Y}=-\frac{2(k_1+k_2)}{m}Y-\frac{k_1k_2 a_0^2}{2m(k_1+k_2)}Y^{\prime\prime}.
\end{equation}
This equation has the form:
\begin{equation} \label{KFG1}
  \ddot{Y}+c^2 Y^{\prime\prime}+\omega^2_0 Y = 0,
\end{equation}
where $\omega^2_0=2(k_1+k_2)/m$. It differs from Klein-Fock-Gordon equation only by sign in front of the spatial derivative. Thus, it can be referred to as the Klein-Fock-Gordon equation in the Euclidean space (EKFG).

For finite systems, we should impose proper boundary conditions. Suppose the chain begins with the first unit cell, so that the system of equations~\eqref{x1} and \eqref{y1} starts from $j=1$. Then for the boundary atom Eq.~\eqref{x1} lacks $-k_2(x_1-y_0)$ term, which requires $x_1=y_0$ at the  boundary. For acoustic mode it gives $\partial_z X=0$ boundary condition. For optical phonons $y_0 \approx -x_0$, and the only possibility to satisfy $x_1=y_0$ is to impose $Y=0$ at the boundary.

Eq.~\eqref{KFG1} is easily generalized to higher dimensions, and we finally write down the continuous boundary value problem which describes the longwavelength optical mode in nonpolar crystals:
\begin{eqnarray}
 \label{KFG2}
  \left(\partial^2_t + C_1 \Delta + C_2 \right)  Y  &=& 0, \\
 Y|_{\partial \Omega} &=& 0, \nonumber
\end{eqnarray}
where $\Delta$ is the Laplace operator, $C_1$ and $C_2$ are some constants, which can be expressed via microscopic parameters, or their values can be taken directly from experiment. Eqs.~\eqref{KFG2} constitute the main result of this Section. Below they will be used for solving the optical phonon eigenproblem in nanoparticles.

\begin{figure}
  \centering
  \includegraphics[width=8cm]{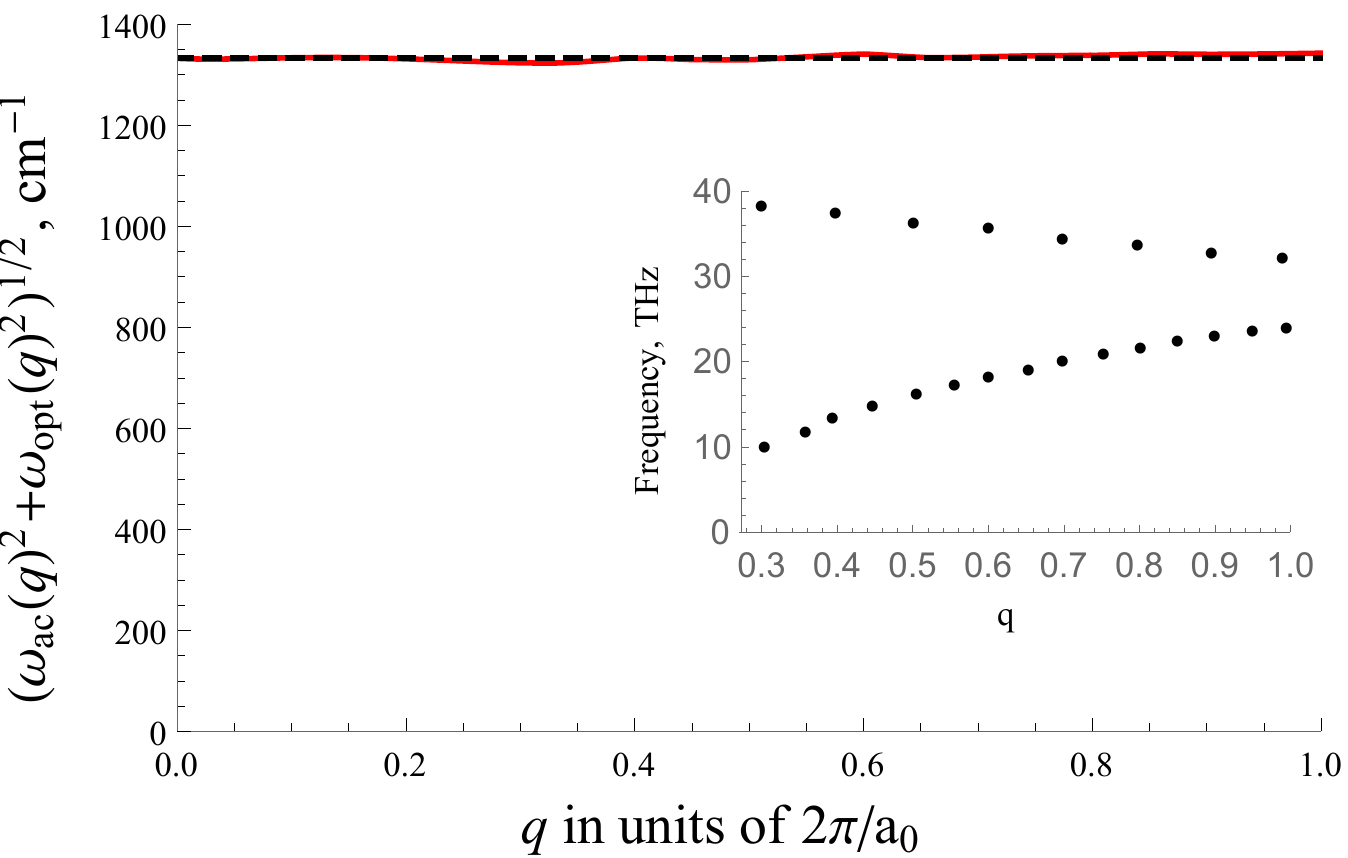}
  \caption{Comparison of experimental value of the quantity $(\omega^2_{ac}(q) + \omega^2_{opt}(q))^{1/2}$ for TA and TO phonons with momenta along the line $\Gamma-X$ in the Brillouin zone extracted from the data of Ref.\cite{monteverde2015}  (see the inset) for bulk diamond  (solid red curve) and its theoretically predicted value $(\omega^2_{ac}(q) + \omega^2_{opt}(q)^{1/2}=\omega_{opt}(0)$ (dashed black curve).  }\label{invar}
\end{figure}

We note, by passing, an interesting feature of the described model:
\begin{equation}\label{Inv1}
  \omega^2_{ac}(q) + \omega^2_{opt}(q) = \omega^2_{opt}(0),
\end{equation}
stemming directly from Eq.~\eqref{Spec1}.  We calculate this relation for transverse acoustic and optical modes in diamond using the experimental data of Ref.~\cite{monteverde2015}. We found that this ``sum rule'' holds in real material, see Fig.~\ref{invar}.

\section{EKFG approach for optical phonons in nanoparticles}
\label{SNano}

The shape of particles depends on the material and on the production method. Obviously, the Laplace-like Eqs.~\eqref{KFG2} can be easily solved numerically on arbitrary manifold. Now we briefly remind to the reader the analytical solutions of these equations for two important cases of sphere and cube.

For certain frequency $\omega$ Eq.~\eqref{KFG2} can be written as follows
\begin{equation} \label{KFG3}
  (\omega^2-C_2)Y-C_1 \Delta Y=0,
\end{equation}
and the corresponding boundary value problem reads
\begin{equation} \label{Laplace}
  \Delta Y + q^2 Y=0,  \qquad Y|_{\partial \Omega} = 0.
\end{equation}
In infinite space it can be solved with plane waves $Y~\propto~\exp(i{\bf qr} - i\omega t)$. The dispersion is given by
\begin{equation}\label{Spec3}
  \omega^2=C_2-C_1 q^2.
\end{equation}
One can rewrite Eq.~\eqref{Spec3} at small momenta $q$ as
\begin{equation}\label{Spec4}
  \omega(q) \approx \sqrt{C_2} - \frac{C_1}{\sqrt{C_2}} \frac{q^2}{2}.
\end{equation}
This is the spectrum with the gap and the negative mass parabolic term. One can relate the spectrum~\eqref{Spec4} originated from EKFG with well-known vibrational models. E.g., within the framework of the Keating \cite{keating1966effect,steiger2011enhanced,anastassakis1990piezo,martin1970elastic,kane1985phonon} model and in agreement with experimental data \cite{warren1967lattice,kulda2002overbending,schwoerer1998phonon,burkel2001determination} the diamond optical phonon dispersion near the Brillouin zone center can be approximated as:
\begin{equation}\label{Keating}
  \omega(q)=A+B \cos{(q a_0/2)} \approx A +B -B \frac{q^2 a^2_0}{8},
\end{equation}
where $a_0$ is the lattice parameter~\cite{note1}. In order to rely Eq.~\eqref{Spec4} to this model we can use either $\{ C_1,C_2 \}$ or $\{A,B \}$ set of parameters. For diamond we adopt the same values of parameters as in Ref.~\cite{koniakhin2018novel}:
\begin{eqnarray}
 \omega_0=  A + B &=& 1333 \, \text{cm}^{-1}, \\
  B &=& 85 \, \text{cm}^{-1}.
\end{eqnarray}

For cubic nanoparticle with edge $a$ one has an obvious standing wave solution normalized to unity and satisfying the boundary condition~\eqref{Laplace}:
\begin{equation}
  \label{WFcub}
  Y_{\bf n}=\sqrt{\frac{8}{a^3}}e^{-i \omega t} \sin{\frac{\pi n_1 x}{a}} \sin{\frac{\pi n_2 y}{a}} \sin{\frac{\pi n_3 z}{a}}.
\end{equation}
Here vector ${\bf n}=(n_1,n_2,n_3)$ enumerates the eigenstates, and the corresponding eigenvalues are
\begin{equation} \label{cubefreqs}
  \omega_{\bf n}=A + B - B \frac{\pi^2}{8} \frac{a^2_0}{a^2}(n^2_1+n^2_2+n^2_3).
\end{equation}
This equation along with Eq.~\eqref{WFcub} will be used below in our RS calculations.

For spherical nanoparticle with radius $R$ Eq.~\eqref{Laplace} obtains the form
\begin{equation}
  \frac{1}{r^2} \frac{\partial}{\partial r} r^2 \frac{\partial Y}{\partial r} - \frac{\hat{l}^2}{r^2} Y + q^2 Y=0,
\end{equation}
where $\hat{l}^2$ is the operator of the squared angular momentum. This equation has the well-known spherical wave solution:
\begin{equation}
\label{eq_spheresolution}
  Y_{\bf n}=R_{ql}(qr)Y_{lm}(\theta, \varphi),
\end{equation}
where $\mathbf{n}=(n,l,m)$ and $Y_{lm}$ are the spherical functions,
\begin{equation}
  R_{ql}(x=qr)=j_l(x)=\sqrt{\frac{\pi}{2x}} J_{l+1/2}(x),
\end{equation}
with $J_{l+1/2}(x)$ being the Bessel functions. The boundary condition requires $J_{l+1/2} (qR)=0$, and the corresponding phonon frequency is given by Eq.~\eqref{Spec3}. The minimal eigenvalue $q^2$ corresponds to $j_0(qR)=\sin(qR)/(qR)=0 $ providing $qR=\pi n$. Optical phonon frequencies for the most important spherically symmetrical modes ($l=m=0$) are given by
\begin{equation}\label{spherefreqs}
  \omega_{n}=A + B - B \frac{\pi^2}{8} \frac{a^2_0}{R^2} n^2.
\end{equation}

\begin{figure}
  \centering
  \includegraphics[width=8cm]{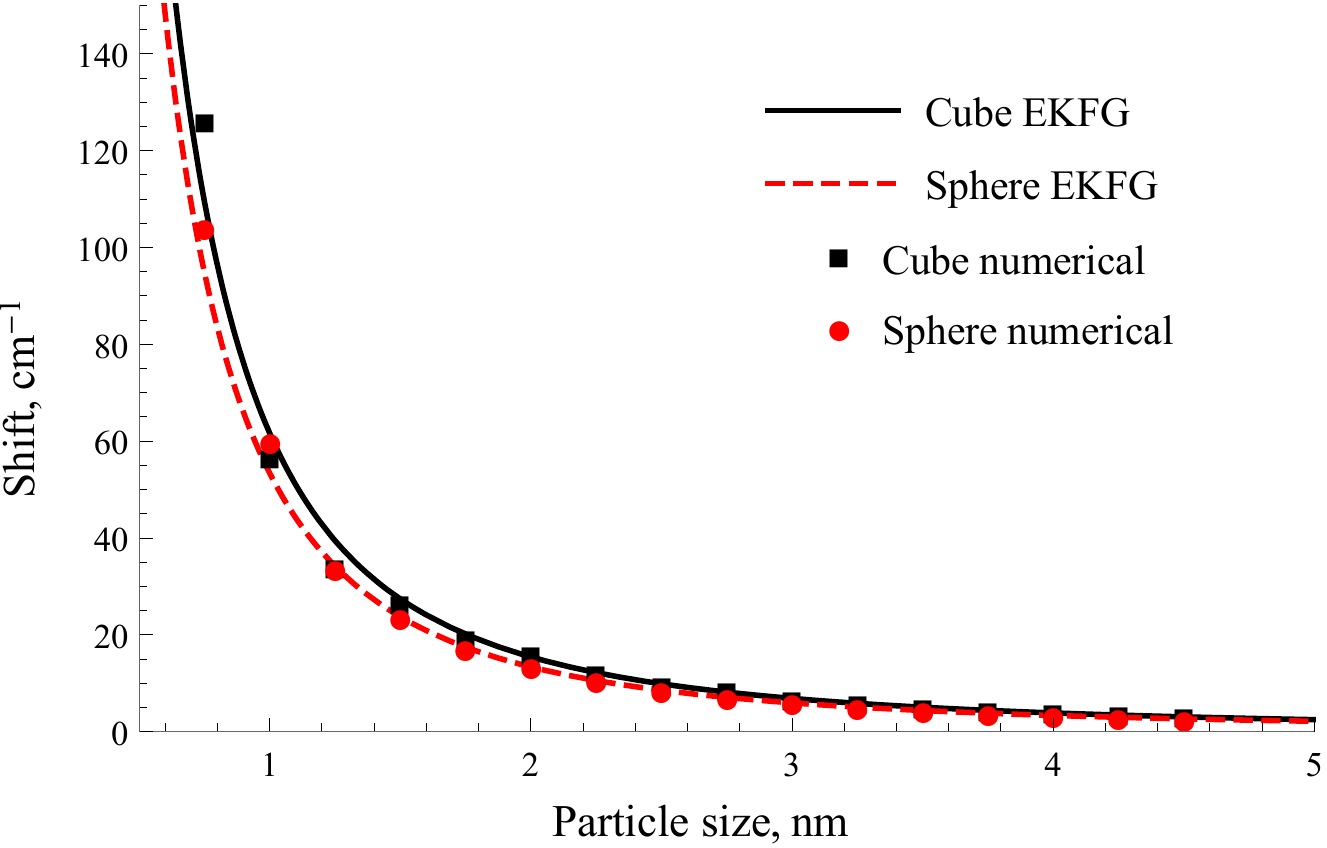}\\
  \caption{The shift of the maximal optical phonon frequency from its bulk value $\omega_0$ as a function of size (see text) calculated using EKFG approach (lines) and obtained numerically within the DMM (dots). Two shapes are considered, cubes (black) and spheres (red). An excellent agreement between analytical approach and numerics is reached for large enough ($ \gtrsim 1$~nm) particles.}\label{shifts}
\end{figure}

\begin{figure}
  \centering
  \includegraphics[width=8cm]{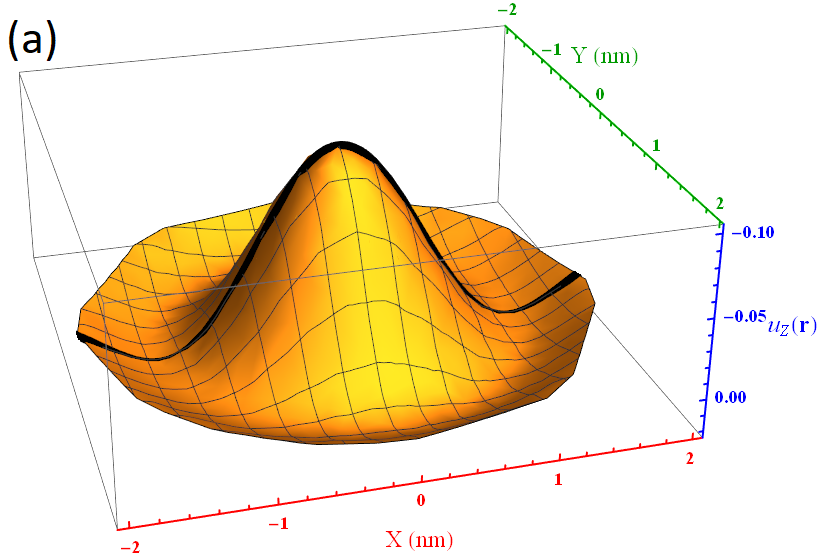}\\
  \includegraphics[width=8cm]{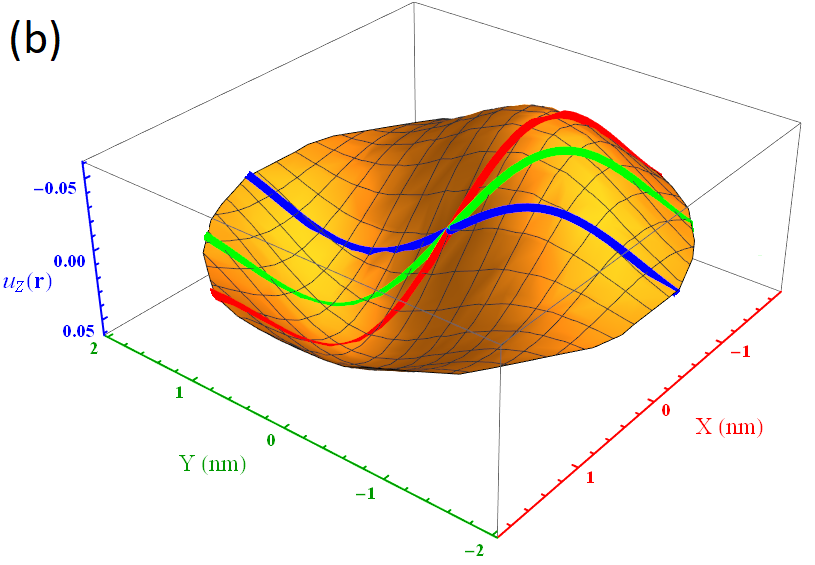}
  \caption{Coinciding spatial structure of eigenfunctions (a)  $Y_{(2,0,0)}$ and (b) $Y_{(1,1,0)}$ of EKFG equation and the envelopes for corresponding eigenfunctions of DMM approach for spherical 4~nm nanodiamond. The surfaces show simultaneously the amplitudes of atomic displacement in the thin section  passing through its center for DMM and the curves show $Y_\mathbf{n}(x,y,0)$ for EKFG.
  \label{envelope}}
\end{figure}


Henceforth, the measure of the \emph{size} for all the particles of arbitrary shape will be the diameter of a sphere with equal volume. For example, the size $L$ of a cubic particle with edge $a$ is
\begin{equation}
  L=\left(\frac{6}{\pi}\right)^{1/3} a.
\end{equation}

In order to verify our approach we compare the size dependence of the maximal optical phonon frequency obtained for nanodiamonds within the framework of EKFG theory and by means of numerical diagonalization of the dynamical matrix, see Fig.~\ref{shifts}. For particles larger than $1$nm our continuous approach fits numerics excellently.

EKGF and DMM approaches yield similar results not only for frequencies of optical phonons. Their similarity goes further. This is illustrated in Fig.~\ref{envelope} where we depict the coincidence of $Y_{(2,0,0)}$ and $Y_{(1,1,0)}$ eigenfunctions obtained within the EKFG and the envelopes of corresponding DMM eigenfunctions calculated both for $4~\text{nm}$ spherical particles.

\section{Continuous bond polarization model}
\label{SRS}

In this Section we show how one can use the solutions of the EKFG eigenproblem for the description of nanoparticle Raman spectra. First, we reformulate the bond polarization model for continuous case. 

According to the BPM~\cite{jorioraman,snoke1993bond,menendez2000vibrational,Guha1996empiricalBPM}, the Raman spectra of nanoparticles and macromolecules can be calculated using the  expression
\begin{equation}\label{BP1}
  I_{\eta^\prime \eta}(\omega) \propto \omega_L \omega^3_S \sum\limits^{3\mathcal{N}}_{f=1} \frac{\langle n(\omega_f) \rangle + 1 }{\omega_f} I_f \delta(\omega-\omega_f),
\end{equation}
where $I_f = \left| \eta^\prime_\alpha \eta_\beta P_{\alpha \beta, f} \right|^2$. In Eq.~\eqref{BP1} $\omega_L$ ($\omega_S$) and $\eta$ ($\eta^\prime$) are the incident (scattered) light frequency and polarization unit vector, respectively, $\langle n(\omega_f) \rangle$ is the $f$-th phonon mode occupation number, $P_{\alpha \beta, f}$ are the polarization tensors of these modes, $\alpha$ and $\beta$ are the Cartesian coordinates. Total nanoparticle polarization tensor is a sum of polarization tensors of each bond in the nanoparticle:  $P_{\alpha \beta, f} = \sum_b P_{\alpha \beta, f}(b)$, where $b$ denote the bond index. The quantities $P_{\alpha \beta, f}(b)$ are linear in the displacements of atoms forming the bond $b$ and corresponding to the phonon mode $f$. After the replacement of summation over bonds $b$ by the volume integration the single mode Raman intensity becomes proportional to the quantity:
\begin{equation}
  \label{Int1}
  I_{\bf n} =  \left| \, \int Y_{\bf n} \, dV  \, \right|^2,
\end{equation}
the factor $\left| \int Y_{\bf n} \, dV \right|^2$ being the analogue of the structure factor in other scattering problems. More precisely, it is the partial structure factor for scattering by the $\mathbf{n}$-th mode.

Substituting Eq.~\eqref{WFcub} to Eq.~\eqref{Int1} we obtain for cubic particle
\begin{equation} \label{Icub}
  I_{\bf n} = \frac{8^3 V}{\pi^6} \left( \frac{ (n_1 \, \text{mod} \, 2)(n_2 \, \text{mod} \, 2)(n_3 \, \text{mod} \,2)}{n_1 n_2 n_3} \right)^2,
\end{equation}
where the symbol $(n_i \, \text{mod} \, 2)$ stands for the remainder of the division of $n_i$ by $2$, and $V$ is the particle volume. 

For spherical particles it is obvious from definition~\eqref{Int1} that the Raman intensity is nonzero only for eigenstates with $l=m=0$. We should properly normalize the eigenfunctions
\begin{equation}
  b^2 \int^R_0 r^2 \left(\frac{\sin{(\pi n r/R)}}{\pi n r/R}\right)^2 dr=1, \qquad b=\frac{\sqrt{2} \pi n}{R^{3/2}}.
\end{equation}
The intensity of these spherically symmetrical modes reads
\begin{equation} \label{Isph}
  I_n = \frac{6V}{(\pi n)^2}.
\end{equation}

Now we are ready to calculate the Raman spectrum of a single particle:
\begin{equation}\label{RS1}
  I_L(\omega) \propto \sum_{\bf n} I_{\bf n} \frac{\Gamma/2}{(\omega-\omega_{\bf n})^2+\Gamma^2/4}.
\end{equation}
Here, the summation runs over all the eigenmodes ${\bf n}$ with corresponding Raman intensities $I_{\bf n}$, $L$ is the particle size, $\Gamma$ is the spectral line broadening, which consists of intrinsic phonon damping and spectrometer resolution. Below this parameter is considered to be the same for any phonon mode. The detailed study of intrinsic mechanisms contributing to $\Gamma$ is out of the scope of the present paper.

It is easy to generalize Eq.~\eqref{RS1} for powders with certain distribution function $n(L)$:
\begin{equation}
\label{eq_Ipowder}
  I(\omega) = \sum_{L} I_L(\omega) n(L).
\end{equation}

\section{Comparison with experiment and other theories}
\label{SCompar}

In this Section we compare the developed EKFG theory with theoretical DMM-BPM and PCM approaches and apply them for interpreting the available experimental data.

Firstly, it is pertinent to remind the general structure of the Raman spectrum as it was obtained within the DMM-BPM theory. The latter was successful in fitting  the experimental data of Refs.~\cite{shenderova2011nitrogen,stehlik2015size,stehlik2016high}. According to Ref.~\cite{koniakhin2018novel} the DMM-BPM spectrum consists of (i) three-fold degenerate first main peak, (ii) the band of ``Raman-silent'' modes (from the 4th to the 12th), (iii) quasicontinuum (consisting of interleaving bands of silent and active modes) beginning with the first band of about 10 dense active modes. The ratio of intensities (first band from quasicontinuum / first main peak) is about $1 / 3$, whereas the contribution of the rest of the spectrum is negligible.


We observe that the Raman spectrum given by EKFG theory contains the single peak in place of the three-fold degenerate peak of DMM-BPM (the difference originates from scalar character of EKFG approach), the band of silent modes and then the second peak (degenerate for some shapes) in place of the Raman active band. The ratio of intensities (second peak / first peak) in EKFG is found to be almost the same as (first band / first peak) of DMM-BPM. The rest of the EKFG spectrum gives the minor contribution as well.

\begin{figure}
  \centering
  \includegraphics[width=8cm]{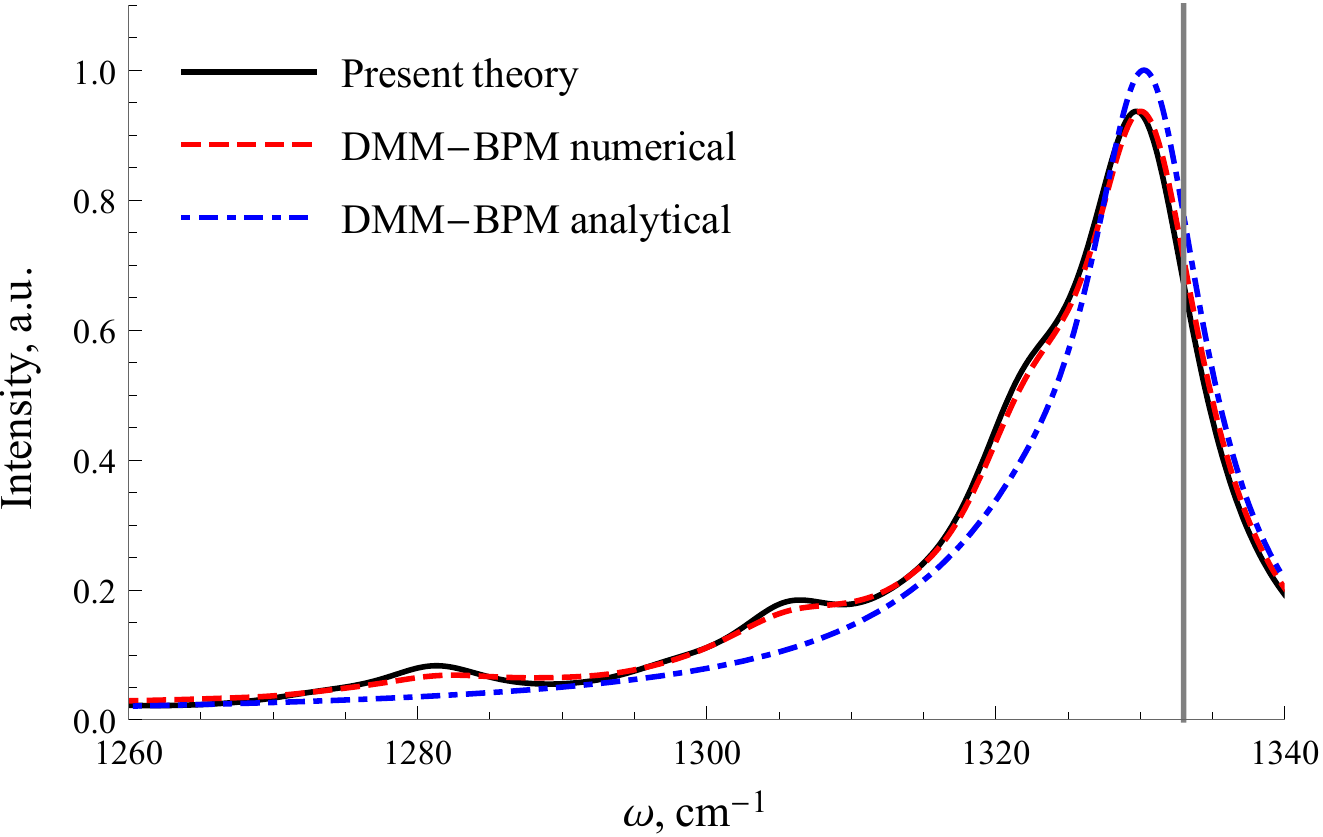}\\
  \caption{Comparison of the RS obtained utilizing EKFG approach and DMM-BPM theory (both numerical and analytical) for 4.5~nm cubic particles, $\Gamma = 10 ~\text{cm}^{-1}$. Grey vertical line demonstrates the position of the bulk diamond peak. }\label{cube45}
\end{figure}

\begin{figure}
  \centering
  \includegraphics[width=8cm]{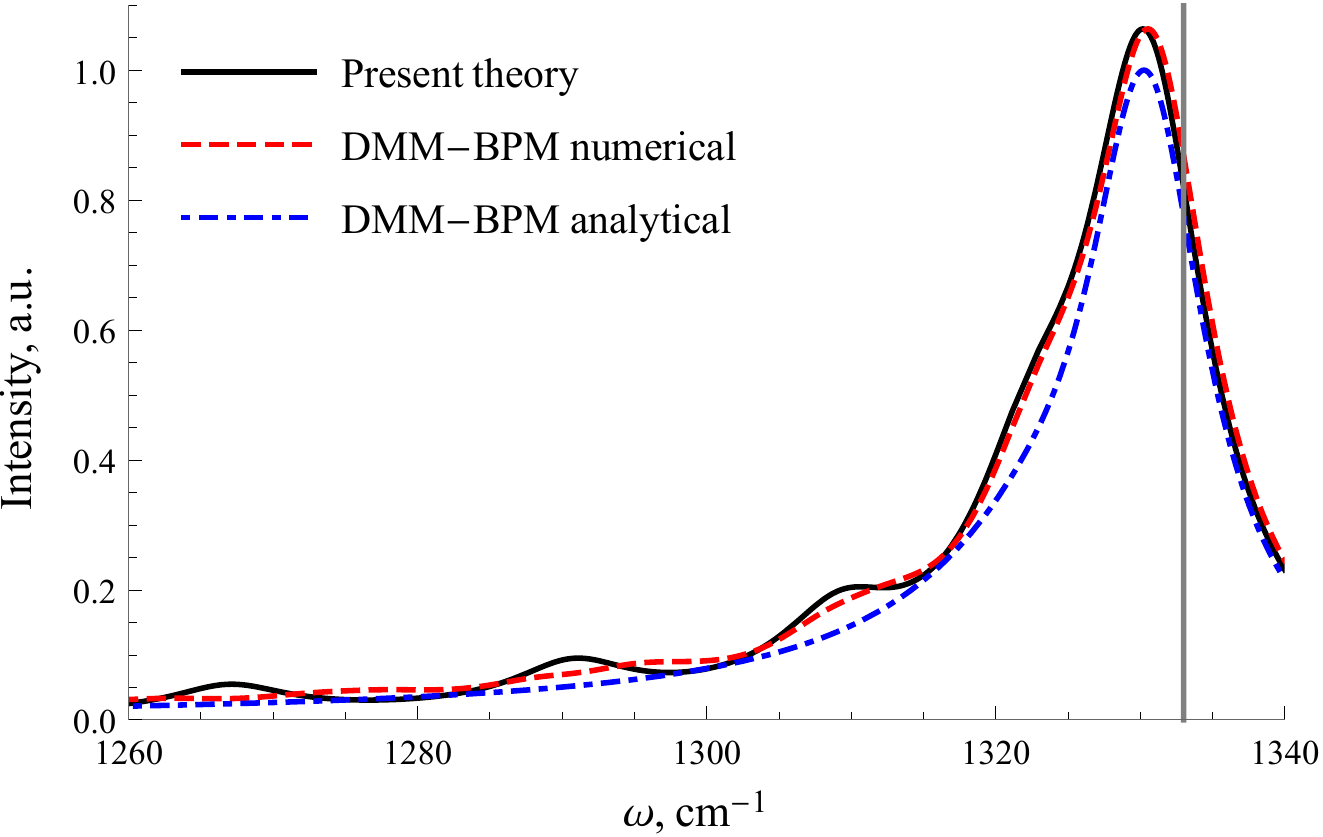}\\
  \caption{Comparison of the RS obtained utilizing EKFG approach and DMM-BPM theory (both numerical and analytical) for 4.5~nm spherical particles, $\Gamma = 10 ~\text{cm}^{-1}$. Grey vertical line demonstrates the position of the bulk diamond peak. }\label{sphere45}
\end{figure}

Furthermore, we observe that the first and the second peaks of EKFG method broadened by experimentally relevant big $\Gamma$'s become almost indistinguishable from broadened first peak and first active band of the DMM-BPM theory, respectively. Together with similarity in eigenfunctions (see Fig.~\ref{envelope}), it justifies the usage of the EKFG theory.

In order to illustrate the above statements we compare the predictions of these two theories. In Fig.~\ref{cube45} and Fig.~\ref{sphere45} we match the Raman spectra of cubic and spherical 4.5~nm particles predicted by EKFG theory with numerical (exact) and analytical versions of DMM-BPM. We report an excellent agreement between numerical DMM-BPM and EKFG approaches. 

As far as the accuracy of the analytical DMM-BPM is concerned we report the good description of the shape and the position of the main peak, whereas the amplitude of the peak as well as the ``left shoulder'' related to quasicontinuum reveals visible discrepancies varying also with particle shape. We attribute these discrepancies to the approximate character of the analytical DMM-BPM, which uses a single shape-averaged density of states. Therefore, it is not capable to describe the shape-dependent details of the band.    

\begin{figure}
  \centering
  \includegraphics[width=8cm]{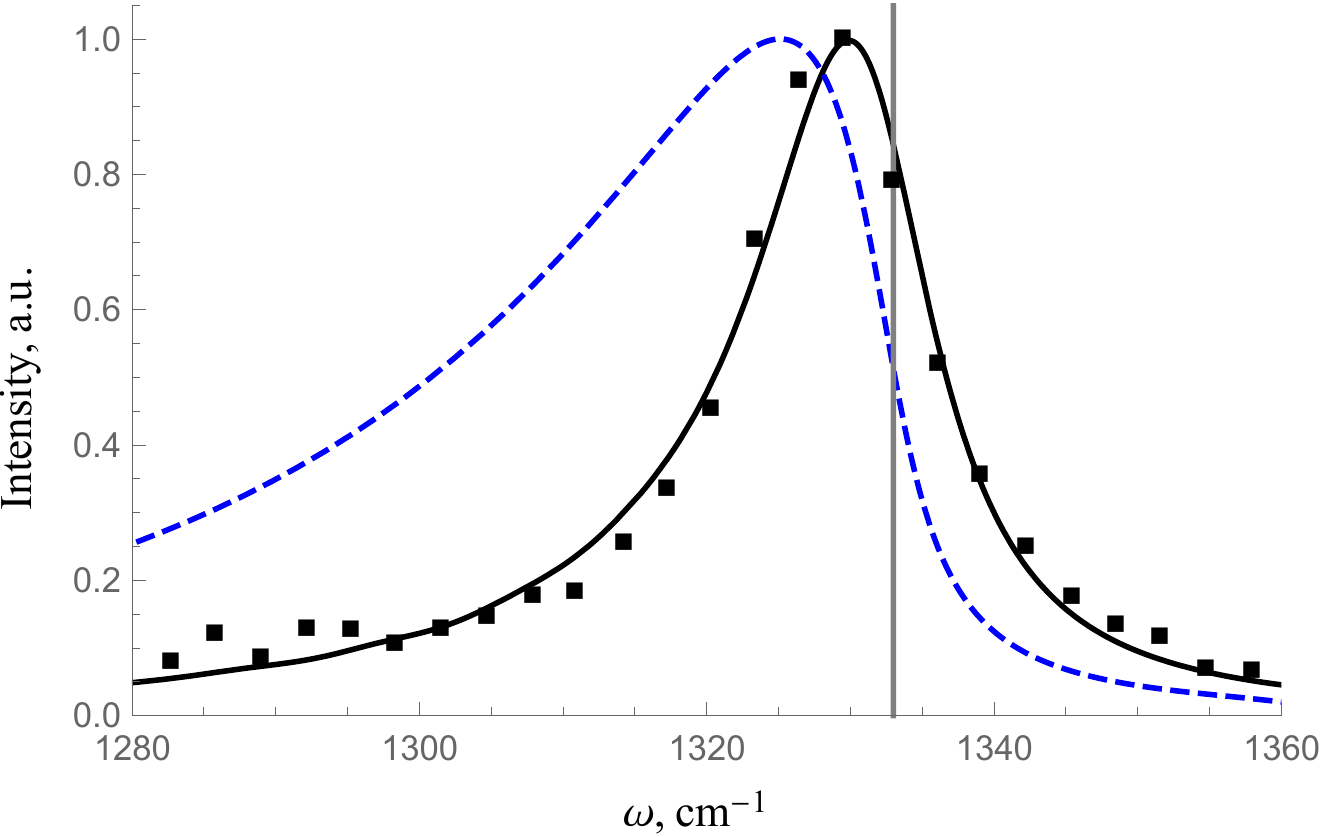}\\
  \caption{Fit of the experimental diamond powder Raman spectrum from Ref.~\cite{stehlik2015size} (dots) within the EKFG approach (black solid line) and within the PCM (blue dashed line). The distribution function is taken from Ref.~\cite{stehlik2015size}. The EKFG calculations are performed for cubic particles and $\Gamma \approx 11.9~\text{cm}^{-1}$. The Ager formulation of the PCM (see Ref.~\cite{ager1991spatially}) is used. Grey vertical line denotes the position of the bulk diamond peak.}\label{ExpDiam}
\end{figure}

\begin{figure}
  \centering
  \includegraphics[width=8cm]{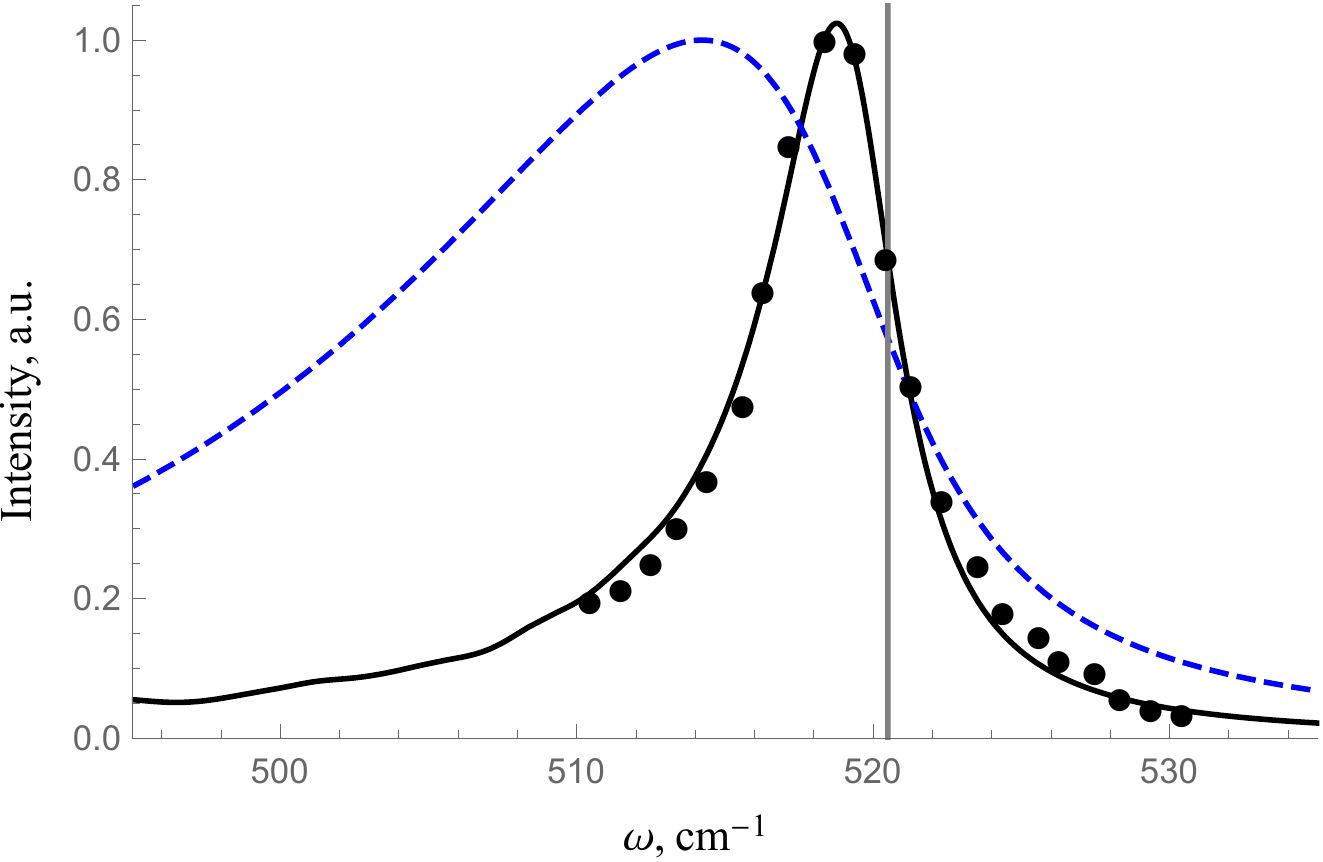}\\
  \caption{Fit of the experimental silicon powder Raman spectrum from Ref.~\cite{expsi2011} (dots) within the EKFG approach (black solid line) and within the PCM (blue dashed line). The distribution function is taken from Ref.~\cite{expsi2011}. The EKFG calculations are performed for spherical particles and $\Gamma \approx 4.2~\text{cm}^{-1}$. The Ager formulation of the PCM (see Ref.~\cite{ager1991spatially}) is used. Grey vertical line denotes the position of the bulk silicon peak.}\label{ExpSi}
\end{figure}

We further proceed with analysis of recent experiment on nanodiamond RS published in Ref.~\cite{stehlik2015size}. We fit this experiment with the use of the distribution function reported in Ref.~\cite{stehlik2015size} and the broadening line parameter $\Gamma \approx 11.9~\text{cm}^{-1}$. The result of our fit is shown in Fig.~\ref{ExpDiam}. Furthermore, in order to demonstrate that our approach is applicable for other nonpolar crystals we analyze the Raman spectrum of Si nanocrystallites experimentally studied in Ref.~\cite{expsi2011}. We utilize the distribution function obtained in Ref.~\cite{expsi2011} by means of TEM. For our fit we accept the textbook values of parameters~\cite{tripathi2007,expsi2011} $A \approx 499.5~\text{cm}^{-1} $ and $B~\approx 21.1 \text{cm}^{-1}$, and find $\Gamma \approx 4.2~\text{cm}^{-1}$. This fit as well as PCM-based calculations are presented in Fig.~\ref{ExpSi}. 

We see that the EKFG curves excellently cover the experimental data. This should be contrasted with the PCM results. Thus, the PCM is able only to \textit{estimate} the particle size, providing for small particles the accuracy of order of hundreds percents.   

The above analysis manifests that the application of EKFG theory for nanoparticle Raman spectra calculations: (i) is successful in fitting the available experimental data, (ii) comparable in accuracy with numerical DMM-BPM and has advantage over the analytical DMM-BPM approach when particle shape is concerned, (iii) works much better than the commonly used PCM.

\section{Shape induced indeterminacy}
\label{SShape}

EKFG theory provides the powerful tool for investigation the Raman spectra dependence on the geometric shape of nanoparticles, namely their faceting. This Section is devoted to this topic. We demonstrate that on the level of accuracy of data analysis presented here the additional information about the particle shape extracted from another type of experiment is desirable for the interpretation. The lack of knowledge about the shape leads to natural limitations in determining the particle size.

\begin{figure}
  \centering
  \includegraphics[width=7cm]{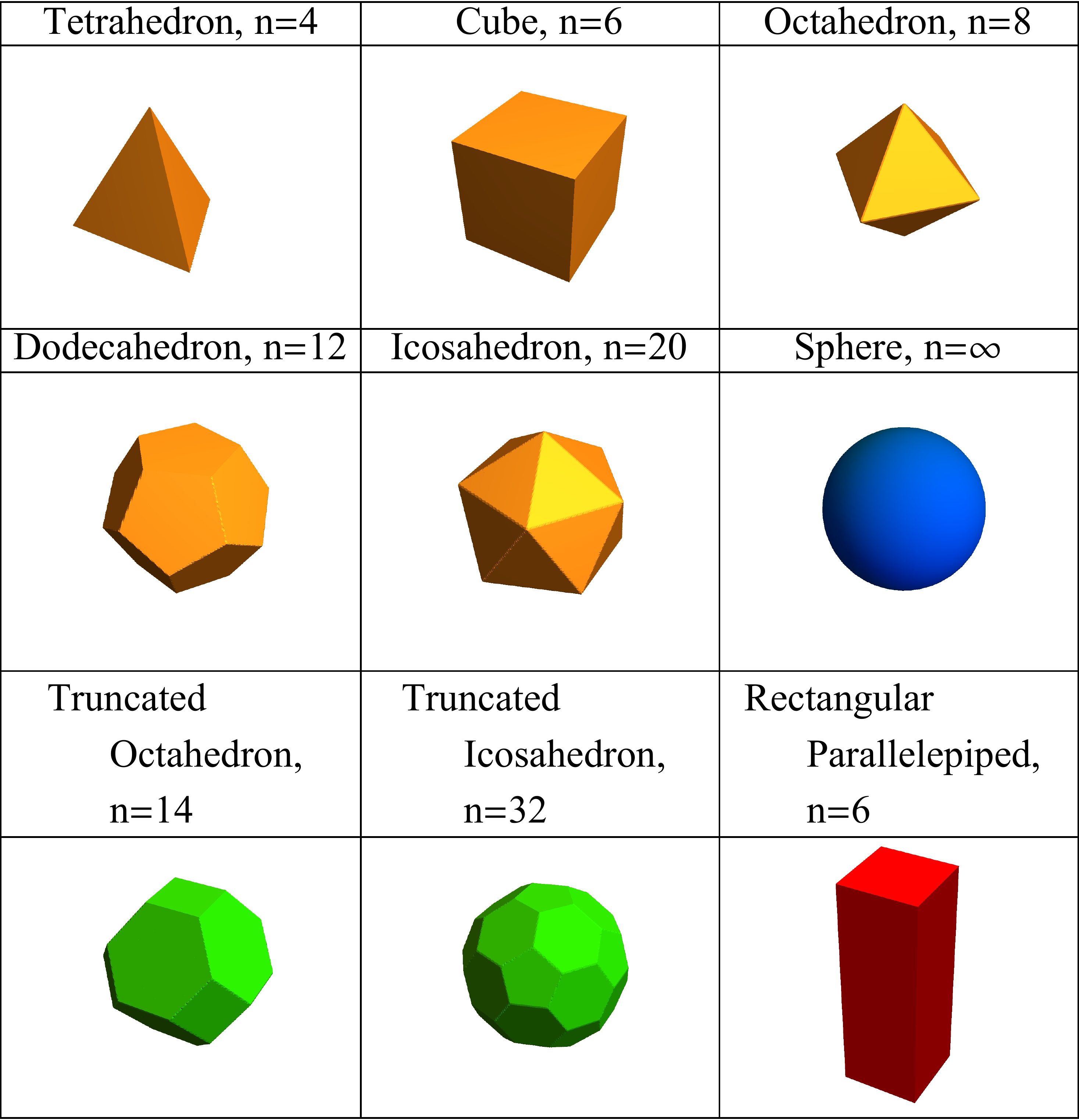}\\
  \caption{Shapes of particles considered in the text. Platonic and Archimedean solids are drawn by orange and green colors, respectively, and the sphere is blue. The rectangular parallelepiped (red) is an example of the elongated solid.} \label{shapes}
\end{figure}

When considering the boundary value problem the shape of the boundary evidently affects both the spectrum and the eigenfunctions of the problem. We investigate this phenomenon by solving the boundary value problem given by Eqs.~\eqref{KFG2} for five Platonic solids and for a sphere as the liming case of Platonic solid with infinitely large number of faces $n$. Specifically (see Fig.~\ref{shapes}), we consider tetrahedron ($n=4$), cube ($n=6$), octahedron ($n=8$), dodecahedron ($n=12$), icosahedron ($n=20$) and sphere ($n=\infty$). In order to study the shapes from wider class we additionally treat two Archimedean solids, namely truncated octahedron ($n=14$) and truncated icosahedron ($n=32$). Furthermore, we address the problem how the elongation of a particle affects its Raman spectrum by studying the particles in shapes of ``prolate'' and ``oblate'' rectangular parallelepipeds ($n=6$) with the edge ratios (3:2:2) and (3:3:2), respectively. The size of particles is chosen to be the same for all shapes.

We observe that the position of the main peak varies with the shape of a particle. For Platonic solids it can be described by simple empirical law:
\begin{equation}\label{Shape1}
  \Delta \omega (n)  \frac{L^2}{a^2_0}   \approx B \left(\frac{\pi^2}{2} + \frac{B_1}{n^2} \right) = 420 \left( 1 + \frac{6.4}{n^2}\right) \text{cm}^{-1},
\end{equation}
where $B_1 = 31.6$. The numbers in the equation above are given for a diamond with $B=85$cm$^{-1}$. Obviously the downshift $\Delta \omega$ is connected with nanoparticles size $L$ and decreases with increasing $L$.

Eq.~\eqref{Shape1} is closely related to the Rayleigh-Faber-Krahn inequality\cite{daners2006faber,faber1923beweis,krahn1925rayleigh,kornhauser1952variational} stating that the first eigenvalue of the problem~\eqref{KFG3} is smallest for a sphere as compared with  other manifolds with equivalent volume. With increasing the facets number the high symmetry polyhedra tend to sphere and we observe that the corresponding eigenvalues monotonically decrease.


Importantly, equation \eqref{Shape1} allows the immediate determination of typical nanoparticle size $L$ in the powder from the Raman peak downshift $\Delta $ with respect to the bulk, cf. with Eqs.~(3) and (7) from Ref.~\cite{koniakhin2018novel} and Fig. 1 from Ref.~\cite{koniakhin2018novel}. See also the discussion after Eq.~(12) in Ref.~\citep{Nemanich1981Light}.

\begin{figure}
  \centering
  \includegraphics[width=8cm]{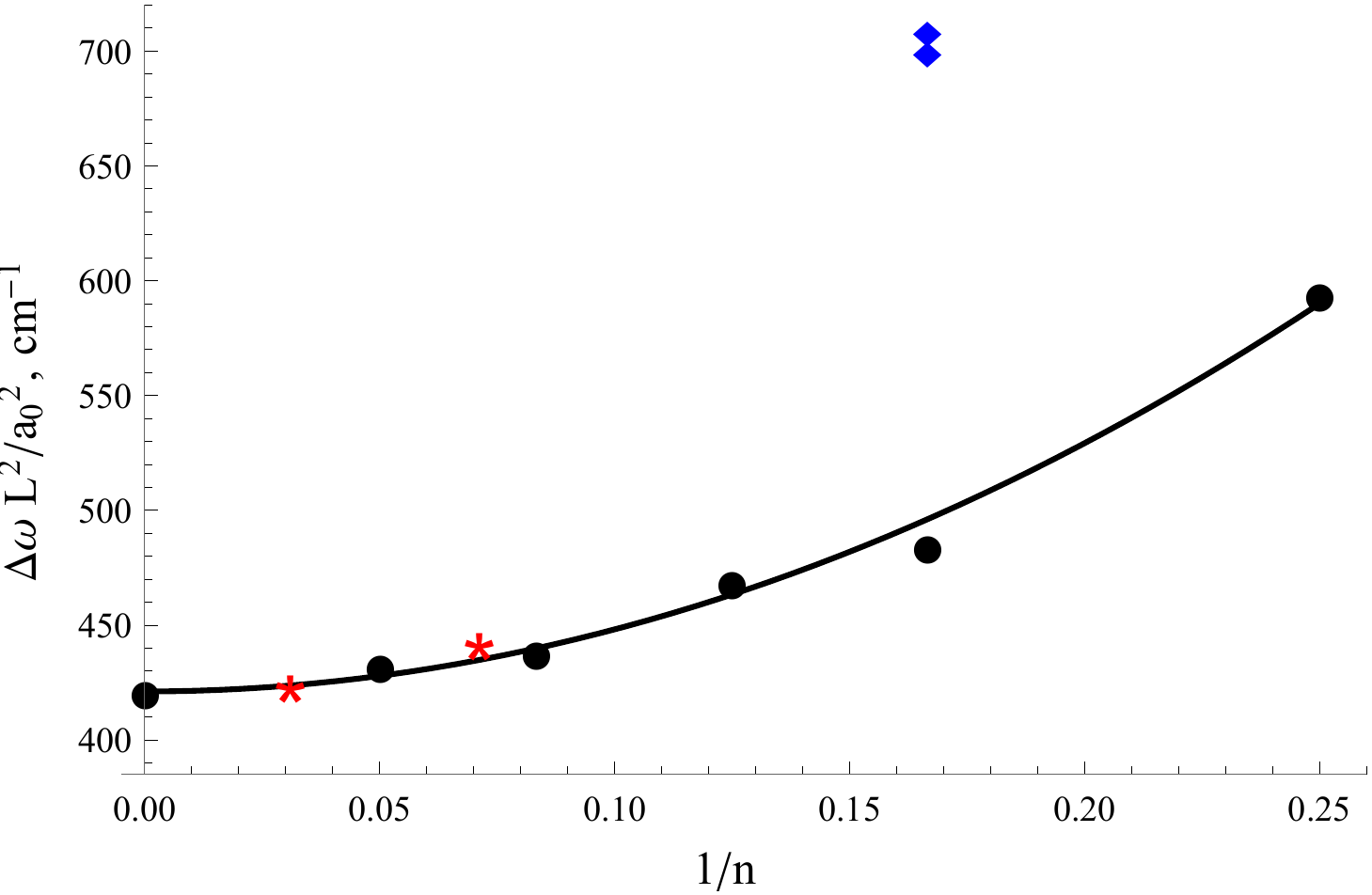}\\
  \caption{Main peak shift as a function of number of faces. Black dots stand for sphere and five Platonic solids. The curve is plotted using the simple ``$1/n^2$'' law (Eq.~\eqref{Shape1}), which nearly holds also for two Archimedean solids (red stars), but is violated for rectangular parallelepipeds (blue diamonds). One can see that the groundless choice of the shape can result in size error up to $18\%$ (see the text).}\label{shapeshift}
\end{figure}

\begin{figure}
  \centering
  \includegraphics[width=8cm]{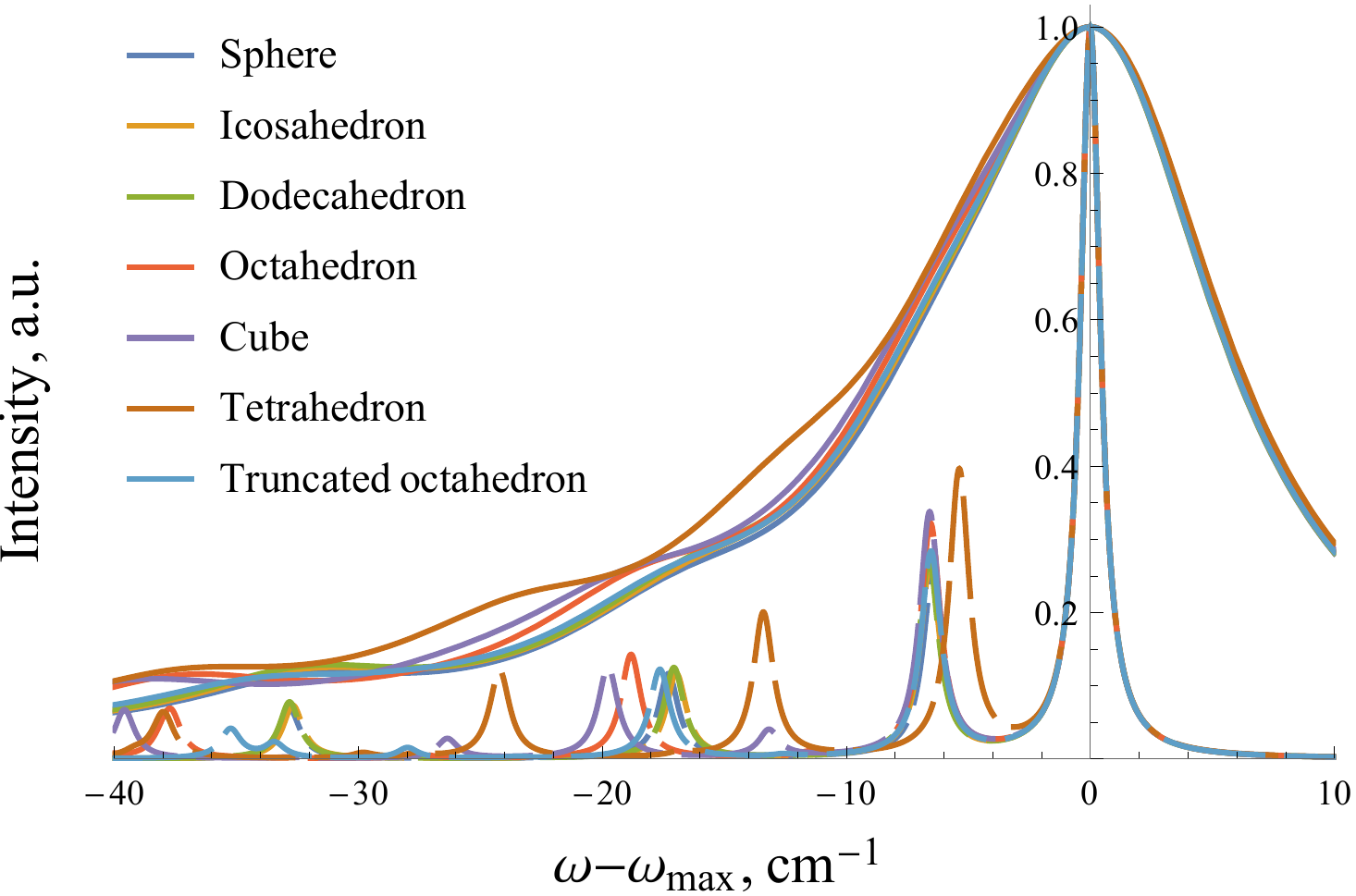}\\
  \caption{Raman spectra for $5~\text{nm}$ diamond particles with different shapes. All the curves are normalized and centered at their maxima (in reality all the maxima differ and lie within the interval $ \sim 1~\text{cm}^{-1}$). At small $\Gamma=1~\text{cm}^{-1}$ (dashed lines) there is some difference in the fine structure of the RS but not of the main peak. In contrast, at $\Gamma=12~\text{cm}^{-1}$ (solid lines) all the curves collapses and obtain almost universal form, the minor differences appearing on the left shoulder are due to higher harmonics.}\label{univ12}
\end{figure}

This dependence is illustrated in Fig.\ref{shapeshift}, 
We also observe that the Archimedean solids nearly belong to the same curve, the truncated octahedron slightly differs from it and the shift is a bit larger then for dodecahedron. We speculate that Eq.~\eqref{Shape1} is approximately valid for a wider than Platonic class of solids. In fact, what is essential is the nearly equal size of a particle along all three spatial dimensions. On the contrary, for elongated particles we observe the drastic deviations from the empirical curve Eq.~\eqref{Shape1} (see Fig.~\ref{shapeshift}). Therefore, the elongated particles should be analyzed separately from the highly symmetrical ones.

In Fig.~\ref{univ12} we plot the RS for all the above shapes of particles, except for the elongated ones and the truncated icosahedron, the latter being indistinguishable from the sphere. We normalize the curves and centered them at their maxima. Although the fine structure of spectra is different for various shapes (as it is seen from Fig.~\ref{univ12} with $\Gamma=1\, \text{cm}^{-1} $), we observe that for realistic value $\Gamma=12\, \text{cm}^{-1} $ the curves generally collapse within the ``main peak approximation'' (some features appear only on the left shoulder).

Therefore, we conclude that the ambiguity in particle shape creates the natural indeterminacy in particle size as it is obtained from the Raman spectrum. It is also seen from Eq.~\eqref{Shape1}. For instance, with no preliminary information about the shape and without an analysis of the shoulder features the value of the particle size extracted from the RS under the assumption that it is a tetrahedron or it is a cube will differ from the case of a sphere by 18\% or 9 \%, respectively. Nevertheless, it is still much better than the PCM can provide.

\section{Raman spectra of powders}
\label{SScale}

In this Section we propose a simple recipe how to calculate the Raman spectrum of a powder with arbitrary size distribution function $g(L)$ starting from the Raman spectrum of a single particle of size $L$ and utilizing the $\Delta \omega(L)$ dependence, see Fig.~\ref{shifts} (cf. Fig.~1 of Ref.~\cite{koniakhin2018novel}).

As it was mentioned before, the main computational difficulty of the numerical DMM-BPM is the cumbersome procedure of $3N \times 3N$ dynamical matrix diagonalization for nanoparticles larger than 5 nm. Thus, the problem is getting almost unsolvable for powders containing large particles, where the dynamical matrix should be diagonalized for each particle size independently. In contrast, the EKFG theory makes the analysis of the powder RS simple and straightforward for any nanoparticles size and their peculiar distribution. This stems from the fact that the eigenvalues of the boundary value problem \eqref{Laplace} obey the scaling law
\begin{equation} \label{Scale1}
  q^2(L_1)=\left( \frac{L_2}{L_1} \right)^2 q^2(L_2), 
\end{equation}
where $L_{1,2}$ are two different particle sizes. Thus, the deviation from the bulk phonon frequency at the Brillouin zone center $\omega_0$ is proportional to $1/L^2$ for every optical phonon mode near the Brillouin zone center. As far as the longwavelength phonons provide the main contribution to the spectrum the scaling law~\eqref{Scale1} obtained within the EKFG can be used for rescaling the eigenfrequencies of the numerical DMM-BPM almost everywhere, except for the smallest particles (cf. Fig.~1 of Ref.~\cite{koniakhin2018novel}).

Thus, we arrive to the following recipe. For powders, we may diagonalize only one dynamical matrix for certain particle size $L$ (e.g., the most probable size of the  distribution). Contributions to RS stemming from other sizes can be obtained by replicating the latter with the use of Eq.~\eqref{Scale1}, the corresponding Raman intensities should be rescaled as
\begin{equation} \label{Scale2}
  I_f(L_1)=\left( \frac{L_1}{L_2} \right)^3 I_f(L_2), 
\end{equation}
cf.~Eqs.~\eqref{Icub} and \eqref{Isph}. Here $f$ enumerates the phonon modes. This scaling procedure allows to calculate the powder RS for large particles and broad distribution functions when the multiple diagonalizations of dynamical matrices become challenging.

\begin{figure}
  \centering
  \includegraphics[width=8cm]{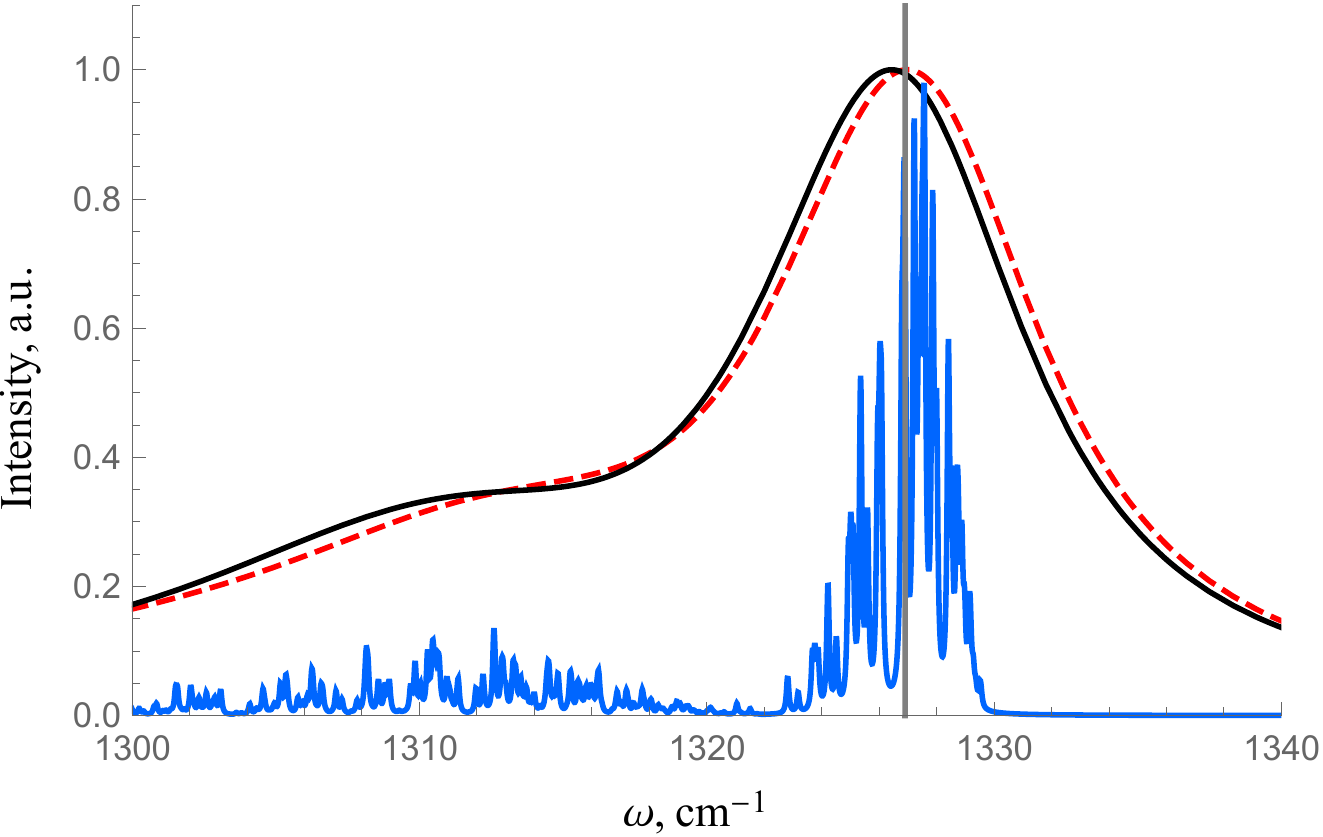}\\
  \caption{ Comparison of the EKFG (solid black curve) and the numerical DMM-BPM (dashed red curve) Raman spectra of a powder, the size distribution is given by Eq.~\eqref{distr1} with $L_0 = 3$~nm and $\sigma=0.33$~nm. The Raman intensities of spectral lines for all particles belonging to distribution \eqref{distr1} are drawn in blue color. The gray line stands for the central peak of the distribution. Notice the downshift of curves maxima as compared to the central peak. }\label{PowderKFG}
\end{figure}

The result of the standard numerical DMM-BPM calculations (cubic particles and no simplifications) for the Gaussian size distribution function 
\begin{equation} \label{distr1}
  g(L) = \frac{1}{2 \pi \sigma} \exp{\left[-(L-L_0)^2/2 \sigma^2 \right]}
\end{equation}
in comparison with the EKFG Raman spectrum smeared out with the use of the aforementioned procedure is presented in Fig.~\ref{PowderKFG}, where $L_0 = 3$~nm and $\sigma=0.33$~nm are taken for both curves. We observe a very good concordance between these two approaches. The linewidth in Fig.~\ref{PowderKFG} is $\Gamma=10~\text{cm}^{-1}$ for both peaks, and the blue lines stand for spectral intensities of all the peaks of different particle sizes $L$ described by the distribution function~\eqref{distr1}. In reality the latter contains step-like contributions stemming from discrete lattice structure. Notice that the maxima of both  broadened curves lie at frequencies lower than the central peak of unbroadened spectrum, while naively one could expect that they coincide. This shift $\sim 10 \% $ is due to admixture of the Raman active band to the main peak. Hence, analyzing only the main peak position provides an additional error in data interpretation. It is of the same order as the shape indeterminacy discussed in previous Section.


\begin{figure}
  \centering
  \includegraphics[width=8cm]{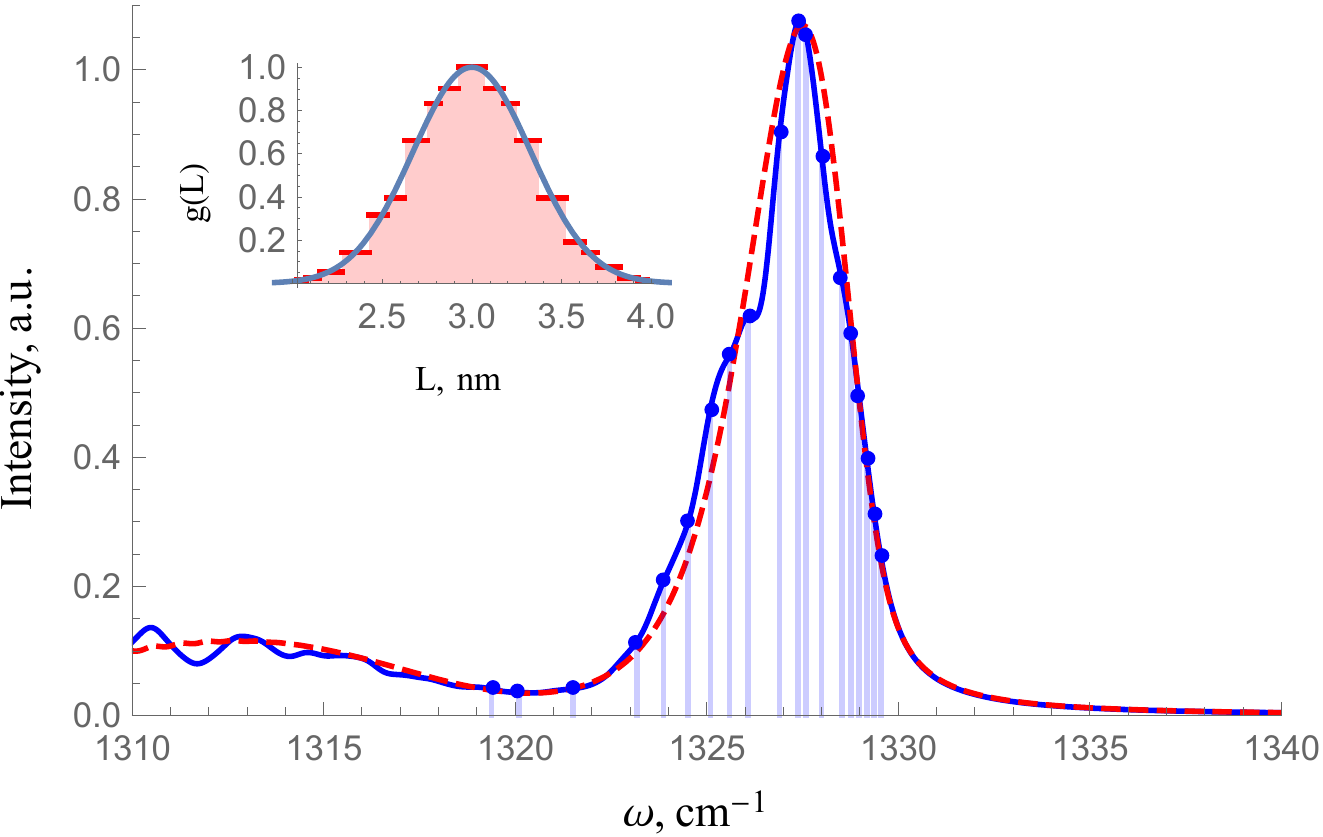}\\
  \caption{Comparison of the standard (solid blue curve)  and rescaled single-particle (dashed red curve) DMM-BPM powder Raman spectra, $\Gamma=1~\text{cm}^{-1}$. Inset shows the distribution function \eqref{distr1} taking into account the lattice induced step-like features. Blue vertical lines describe the positions of highest phonon modes contributing to the Raman peak.  }\label{PowderScaling} 
\end{figure}

Rescaling can be also performed to derive the spectrum $I_{L_2}(\omega)$ of a particle with size $L_2$ from the theoretical spectrum $I_{L_1}(\omega)$ for a particle with  size $L_1$ by means of a formula
\begin{equation}
\label{eq_skscale}
I_{L_2}(\omega)=\left(\frac{L_2}{L_1} \right)^3 \cdot I_{L_1} \left(\omega_0 - (\omega_0-\omega) \left(\frac{L_2}{L_1} \right)^2 \right),
\end{equation}
As an initial $I_{L_1}(\omega)$ one can use the data given in Fig.~\ref{univ12} or in Fig.~5 in Ref.~\cite{koniakhin2018novel}. The applicability conditions are small $\Gamma$ in the original spectrum and narrow size distribution of the powder.

In Fig.~\ref{PowderScaling} we present the powder Raman spectrum of cubic particles distributed in line with Eq.~\eqref{distr1} and calculated within the framework of the standard numerical DMM-BPM scheme in comparison with the smeared out according to the rescaling recipe single-particle DMM-BPM spectrum. The broadening parameter $\Gamma=1~\text{cm}^{-1}$ is chosen unphysically small in order to demonstrate the features caused by discrete character of the distribution function.

Thus, in this Section we proposed the simple ``rescaling'' method to calculate the powder Raman spectra.

\section{Discussion and conclusions}
\label{SSum}

Motivated by poor capability of the conventional phonon confinement model to describe the Raman spectra of very small particles we developed the effective continuous theory of confined optical phonons in terms of the Dirichlet boundary value problem for Klein-Fock-Gordon equation in Euclidean space. Being supplemented with continuous version of the bond polarization model the EKFG approach allows us to obtain the Raman spectra of nanoparticles using the linewidth $\Gamma$ as a single adjustable parameter of the theory.

Our approach developed in Sec.~\ref{SKFG} could be regarded as a simple continuous scalar version of elasticity theory capable to describe both acoustic and optical waves.
In its present formulation EKFG approach is completely sufficient for explaining the nanoparticles Raman spectra with high accuracy. It would be extremely interesting to develop the anisotropic version of this theory for tensor $X$ and $Y$ fields (see Eqs.~\eqref{X2} and \eqref{Y2}) allowing to naturally introduce the notion of polarization for optical phonons.  

We undertook the comparative analysis of two recent experiments on Raman scattering in diamond~\cite{stehlik2015size} and silicon~\cite{expsi2011} powders with precisely measured nanoparticles size distributions using PCM and EKFG approaches in attempt to fit the data. We observe that for such small nanocrystallites the PCM yields an error of the order of $100\%$. On the contrary, the EKFG theory excellently describes the experiments. Moreover, the Raman spectra calculated within the EKFG approach are indistinguishable from those obtained with the use of the microscopic numerical DMM-BPM. As far as the analytical DMM-BPM is concerned the EKFG spectra looks better due to certain simplifications of the former, where in particular the unified shape-independent density of states is utilized.   

Now let us discuss the ranges of employment for all three theories (PCM, DMM-BPM and EKFG) at hands. For particle sizes $L \gtrsim 10~\text{nm}$ the DMM-BPM calculations are cumbersome, PCM works reasonably good and we can use either the EKFG or analytical DMM-BPM or the PCM depending on our choice. For $ 10~\text{nm} \gtrsim L \gtrsim 5~\text{nm}$ the DMM-BPM method is still tedious, whereas the PCM becomes inaccurate, so the best options are the EKFG or analytical DMM-BPM approaches. For smallest particles $ 5~\text{nm} \gtrsim L \gtrsim 1~\text{nm}$ the microscopic DMM-BPM calculations may be performed during reasonable machine time. In most complicated cases they can be accompanied by the express EKFG analysis. Still, the PCM is out of the game.

EKFG and DMM-BPM both provide the description of the same optical phonon size quantization effect in nanoparticles, first being the continuous model and second being the discrete one. Therefore they give similar results and both are more preferable than the PCM when interpreting the experimental data.

Also, our research reveals specific particle shape dependence of the Raman spectra. Therefore, the additional information about this shape from other types of experiment is highly desirable. Otherwise, the lack of this knowledge imposes a limitation on the accuracy of size determining from Raman experiment at the level $\sim 10\%$.

At last, the analysis of scaling properties of the continuous EKFG model allows us to formulate the theoretical method of constructing the Raman spectrum of a powder from  the single-particle Raman spectrum and size distribution function without spectrum calculations for each size. Rigorously, the distribution function in EKFG is continuous; however, it might be useful to utilize the realistic discrete distribution function appropriate for DMM-BPM.  

Throughout this paper as well as in our previous study~\cite{koniakhin2018novel} we treated the Raman spectra appealing mostly to the \textit{main} peak peculiarities. On the other hand, our analysis of shape-dependent properties and powder spectra manifests the importance of the left shoulder features, which are evidently associated with the second line in EKFG and/or with the beginning of the quasicontinuum in DMM-BPM. Although the weight of the second peak is several times smaller than the first one, it turns out that it is more affected by the shape. This opens up the problem of the \textit{lineshape} in Raman spectra. Along with intrinsic broadening of the spectral line this topics are of further interest~\cite{OurPhononDamping}.

We summarize the results of this by presenting the step-by-step recipe how to calculate the Raman spectra of nanoparticles or their powders: 

\begin{enumerate}
\item Choose the geometric shape of particles; solve the eigenproblem Eq.\eqref{Laplace} and find the set of eigenfunctions $Y_{\bf n}$. Eigenfunctions for sphere are given by Eq. \eqref{eq_spheresolution} and for cube by Eq. \eqref{WFcub}.
\item For each mode $\bf n$ calculate its frequency $\omega_{\bf n}$ using Eq.~ \eqref{Keating} (see Eq.~\eqref{spherefreqs} for sphere and Eq.~\eqref{cubefreqs} for cube).
\item For each mode $\bf n$ calculate the Raman intensity using Eq.~\eqref{Int1} (see  Eq. \eqref{Isph} and Eq.~\eqref{Icub}).
\item Perform the rescaling for other sizes using Eqs.~\eqref{Scale1} and \eqref{Scale2}. Or perform steps 2 and 3 for all sizes.
\item Calculate the spectra for particles of all sizes with Eq.~\eqref{RS1}.
\item Account for the size distribution function using Eq.~\eqref{eq_Ipowder} and obtain a spectrum of a powder. 
\end{enumerate}

In conclusion, we propose the novel method to calculate the Raman spectra of nonpolar nanoparticles. We formulate the continuous boundary value problem for Klein-Fock-Gordon equation in Euclidean space with the Dirichlet boundary conditions. We rewrite the bond polarization model in the continuous form and calculate the nanoparticle Raman spectra. 
Our model is shown to fit the recent experimental data excellently. The correspondence of this model with the DMM-BPM approach as well as the failure of the PCM are demonstrated. The role of the particle shape is investigated and the limitations on the accuracy of the method are discussed. The simple recipe to construct the Raman spectra of powders is proposed.

\begin{acknowledgments}

O.I.U. thanks for financial support the Skolkovo Foundation (grant agreement for Russian educational and scientific organisation no.7 dd. 19.12.2017) and Skolkovo Institute of Science and Technology (General agreement no. 3663-MRA dd. 25.12.2017). The contribution to the study conducted by S.V.K. was funded by RFBR according to the research project 18-32-00069. We acknowledge the support of the project "Quantum Fluids of Light" (ANR-16-CE30-0021). Thanks to D. Stupin for discussions and to G. Malpuech for his support.

\end{acknowledgments}
\appendix

\bibliography{KFG}

\begin{thebibliography}{48}
\bibitem [{not({\natexlab{b}})}]{note1}%
  \BibitemOpen
  \href@noop {} {} \bibinfo {note} {Typically, the phonon
  dispersion is written as $\omega(q)=A+B \cos{(\pi\tilde{q})}$, where
  $\tilde{q}$ is the normalized phonon wave vector: $\tilde{q} = 1$ at the BZ
  boundary ($q_{max}\approx\frac{2\pi}{a_0}$).}\BibitemShut {Stop}%
\end{thebibliography}

\end{document}